\documentclass[floatfix,aps,prl,reprint,groupedaddress]{revtex4-2}

\usepackage{graphicx}% Include figure files
\usepackage{dcolumn}% Align table columns on decimal point
\usepackage{bm}% bold math
\usepackage{braket}
\usepackage{csquotes}
\usepackage{amsmath}
\usepackage{amssymb}
\usepackage[separate-uncertainty=true]{siunitx}
\usepackage{xcolor}

\begin{document}

\title{Coherent Interaction of 2s and 1s Exciton States\\in Transition-Metal Dichalcogenide Monolayers}

\author{Max Wegerhoff}
\email{Contact author: max.wegerhoff@uni-bonn.de}
\author{Moritz Scharfstädt}
\author{Stefan Linden}
\author{Andrea Bergschneider}
\affiliation{Physikalisches Institut, University of Bonn, 53115 Bonn, Germany}

%%%%%%%%%%%%%%%%%%%%%%%%%%%%%%%%%%%%%%%%%%%%%%%%%%%%%%%%%%%%%%%%%%%%%%%%%%%%%
%Abstract
%%%%%%%%%%%%%%%%%%%%%%%%%%%%%%%%%%%%%%%%%%%%%%%%%%%%%%%%%%%%%%%%%%%%%%%%%%%%%
\begin{abstract}
We use femtosecond pump-probe spectroscopy to study the coherent interaction of excited exciton states in WSe$_2$ and MoSe$_2$ monolayers via the optical Stark effect. For co-circularly polarized pump and probe, we measure a blueshift which points to a repulsive interaction between the 2s and 1s exciton states. 
The determined 2s-1s interaction strength is on par with that of the 1s-1s, in agreement with the semiconductor Bloch equations. 
Furthermore, we demonstrate the existence of a 2s-1s biexciton bound state in the cross-circular configuration in both materials and determine their binding energy. 
\end{abstract}

\maketitle

%%%%%%%%%%%%%%%%%%%%%%%%%%%%%%%%%%%%%%%%%%%%%%%%%%%%%%%%%%%%%%%%%%%%%%%%%%%%%
%Introduction: general
%%%%%%%%%%%%%%%%%%%%%%%%%%%%%%%%%%%%%%%%%%%%%%%%%%%%%%%%%%%%%%%%%%%%%%%%%%%%%
\textit{Introduction.}---Transition metal dichalcogenide (TMDC) monolayers (ML) feature tightly bound excitons \cite{mak_atomically_2010,splendiani_emerging_2010, wang_colloquium_2018} offering an ideal platform to investigate many-body interactions \cite{sie_intervalley_2015,barachati_interacting_2018, tan_interacting_2020, wei_charged_2023}. These interactions are repulsive for excitons within the same valley, while excitons in opposite valleys interact attractively leading to the formation of biexcitons \cite{hao_neutral_2017, ye_efficient_2018,barbone_charge-tuneable_2018,li_revealing_2018,chen_coulomb-bound_2018,steinhoff_biexciton_2018}.
The strong binding also facilitates the presence of excited exciton states, so called Rydberg excitons \cite{chernikov_exciton_2014, he_tightly_2014}. The latter feature enhanced interactions and are a promising candidate for the realization of optical nonlinearities \cite{walther_giant_2018,gu_enhanced_2021,heckotter_asymmetric_2021} and quantum sensing applications \cite{popert_optical_2022}. 
For this purpose, it is essential to understand and quantify the interaction between Rydberg excitons. 
One possibility is the measurement of the interaction-induced blueshift in a pump-probe scheme.
However, resonant excitation schemes are plagued by the immediate population of dark states making the extraction of the interaction strength unreliable.
To solve this issue, the optical Stark effect has recently been employed to characterize repulsive exciton-exciton interactions \cite{uto_interaction-induced_2024, evrard_ac_2025} as well as the formation of biexcitons due to attractive exciton-exciton interactions \cite{sie_observation_2016,yong_biexcitonic_2018}.

The optical Stark effect is a fundamental concept in atomic physics and quantum optics and refers to the energy shift in a two-level system caused by a non-resonant pump field. This effect also occurs in semiconductors with enhanced light-matter interaction such as quantum wells \cite{mysyrowicz_dressed_1986,von_lehmen_optical_1986}, quantum dots \cite{unold_optical_2004} and TMDC monolayers \cite{uto_interaction-induced_2024,evrard_ac_2025,kim_ultrafast_2014,sie_valley-selective_2015,sie_observation_2016,yong_biexcitonic_2018,cunningham_resonant_2019,slobodeniuk_ultrafast_2023,choi_ultrafast_2024}. The pump interacts coherently with the excitonic states in the material and causes an energy shift of the latter. For large detunings, the energy shift behaves analogously to a two-level system. However, if the detuning becomes comparable to the exciton binding energy, many-body effects begin to play a dominant role \cite{schmitt-rink_collective_1986,zimmermann_dynamical_1988,schmitt-rink_linear_1989,combescot_semiconductors_1992}.

In this work, we use the excitonic optical Stark effect to investigate the exciton-exciton interaction beyond the 1s state in the two archetypal TMDC monolayers WSe$_2$ and MoSe$_2$. The two materials differ significantly in the exciton Bohr radius as well as in the arrangement of the conduction bands. To achieve high comparability between the two materials, we embed them in a charge-controlled heterostructure with nearly identical layer thicknesses and perform the optical measurements at cryogenic temperatures. In addition to the 1s-1s interaction strength, we determine the 2s-1s interaction strength and observe the signature of a 2s-1s biexciton whose binding energy we determine.

%%%%%%%%%%%%%%%%%%%%%%%%%%%%%%%%%%%%%%%%%%%%%%%%%%%%%%%%%%%%%%%%%%%%%%%%%%%%%
%Introduction: Optical Stark effect
%%%%%%%%%%%%%%%%%%%%%%%%%%%%%%%%%%%%%%%%%%%%%%%%%%%%%%%%%%%%%%%%%%%%%%%%%%%%%
We investigate the optical Stark shift using an ultrafast pump-probe setup. The pump pulse is always red-detuned relative to the 1s state, so that no real excitation of charge carriers can take place. Instead the pump field induces a polarization in the semiconductor material that interacts with the real exciton states and leads to an instantaneous energy shift of the latter. This shift is then measured by a small test excitation generated by the probe pulse.
The expected shift contains a contribution from the exciton-photon (XP) interaction as well as the exciton-exciton (XX) interaction \cite{schmitt-rink_linear_1989, haug_quantum_2005}. In the framework of the semiconductor Bloch equations (SBE) \cite{haug_quantum_2005}, the shift $\Delta_\lambda$ for a given exciton state $\lambda$ can be expressed in linear order of the pump intensity $I$ as
\begin{equation}\label{eq:shift_copol}
\Delta_\lambda / I 
= 
\sum_{\lambda^\prime}
\left(
a_{\lambda \lambda^\prime} \frac{1}{\delta_{\lambda^\prime}}
+ b_{\lambda \lambda^\prime} \frac{1}{\delta_{\lambda^\prime}^2}
\right),
%\quad, \quad \lambda^\prime = 1\text{s},2\text{s},  \dots
\end{equation}
where the contribution of all possible excitonic states $\lambda^\prime$ are summed up and $\delta_{\lambda^\prime}=E_{\lambda^\prime}- E_{\text{pump}}$ denotes the detuning between the respective excitonic transition and the pump.

The contribution of each state $\lambda^\prime$ to the polarization in the material is proportional to $1/\delta_{\lambda^\prime}$. This simultaneously results in a virtual exciton density $n_{\lambda^\prime}$, which  is given by the square of the polarization and therefore scales as $1/\delta_{\lambda^\prime}^2$ \cite{zimmermann_dynamical_1988,schmitt-rink_linear_1989,haug_quantum_2005}.
For the XP interaction, the polarization induces a dipole moment for the state $\lambda$. The pump interacts with this dipole moment and causes an energy shift.
In the case of the XX term, the virtual density interacts with the state $\lambda$ with an interaction strength $U_{\lambda \lambda^\prime}$ causing an additional energy shift $n_{\lambda^\prime}\, U_{\lambda \lambda^\prime}$ \cite{schmitt-rink_linear_1989, haug_quantum_2005, cunningham_resonant_2019, uto_interaction-induced_2024}.
The parameters $a_{\lambda \lambda^\prime}$ subsume all prefactors for the XP terms including the induced polarizations, while the parameters $b_{\lambda \lambda^\prime}$ absorb the prefactors of the XX terms including the induced densities and interaction strengths \cite{SM}. By analyzing the light shift for different detunings, we disentangle the XP and XX contributions as well as the contribution of the different exciton states.

%%%%%%%%%%%%%%%%%%%%%%%%%%%%%%%%%%%%%%%%%%%%%%%%%%%%%%%%%%%%%%%%%%%%%%%%%%%%%
%Figure 1: Sample, Micrograph, gated DRS
%%%%%%%%%%%%%%%%%%%%%%%%%%%%%%%%%%%%%%%%%%%%%%%%%%%%%%%%%%%%%%%%%%%%%%%%%%%%%
\begin{figure}
\includegraphics{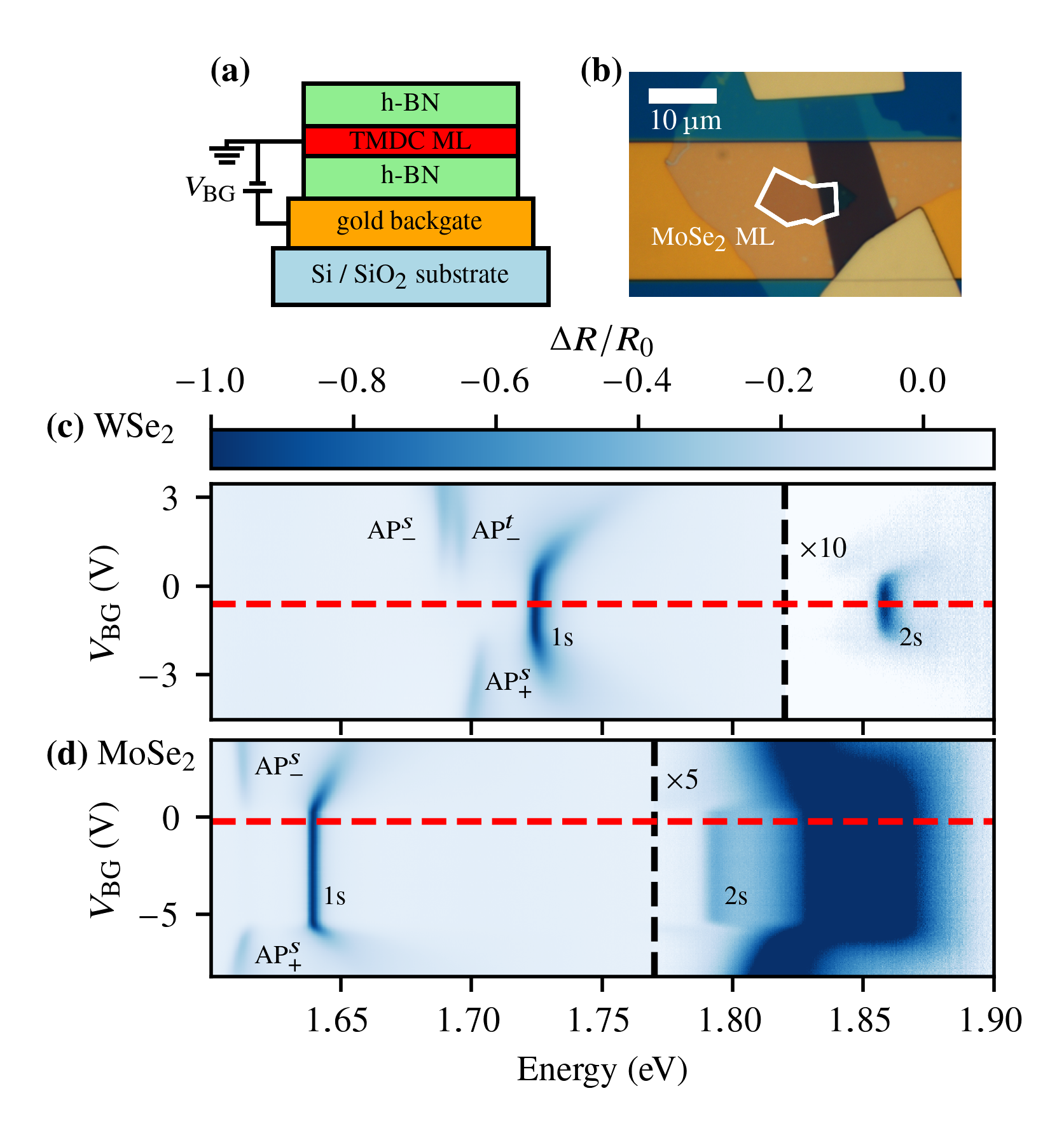}
\caption{\label{fig1} (a) Schematic of the van-der-Waals heterostructure used for both devices. (b) Optical micrograph of the MoSe$_2$ device. The electrical contact to the ML is established by a graphite flake. (c),(d) Differential reflection spectra of the WSe$_2$ and MoSe$_2$ device as a function of the backgate voltage $V_\text{BG}$. In (c) the signal is multiplied by $10$ for $E>\SI{1.82}{\electronvolt}$, and in (d) by $5$ for $E>\SI{1.77}{\electronvolt}$ to enhance the visibility of the 2s resonance. The red dashed lines mark the voltages in the charge-neutral region at which the pump-probe measurements were carried out. With increasing doping, the 1s and 2s states evolve into the respective repulsive polarons (RP). At the same time, the attractive polaron (AP) branches emerge \cite{sidler_fermi_2017,efimkin_many-body_2017,liu_exciton-polaron_2021,huang_quantum_2023}.}
\end{figure}
%%%%%%%%%%%%%%%%%%%%%%%%%%%%%%%%%%%%%%%%%%%%%%%%%%%%%%%%%%%%%%%%%%%%%%%%%%%%%
%Sample and setup
%%%%%%%%%%%%%%%%%%%%%%%%%%%%%%%%%%%%%%%%%%%%%%%%%%%%%%%%%%%%%%%%%%%%%%%%%%%%%
\textit{Sample and setup.}---We perform measurements on two different devices (Fig.\,\ref{fig1}(a)) with either a WSe$_2$ or MoSe$_2$ monolayer encapsulated in h-BN and control the doping of the monolayer with a gold backgate. Fig.\,\ref{fig1}(b) depicts an optical micrograph of the MoSe$_2$ sample. The differential reflection spectra of both samples (Fig.\,\ref{fig1}(c) and (d)) show the 1s and 2s excitonic states at charge neutrality at which the pump-probe measurements are carried out (see red dashed line). 
Due to the combination of a thick gold backgate and thin h-BN layers (each $\leq\SI{20}{nm}$), the absorption by each exciton transition leads to a simple Lorentzian dip in the reflection spectra with line widths of around $\SI{4}{\milli\electronvolt}$ for the 1s and $\SI{6}{\milli\electronvolt}$ for the 2s transition. The bottom h-BN thickness is large enough such that screening effects due to the gold backgate are negligible \cite{tebbe_tailoring_2023}.

The measurements of optical Stark shift are performed inside a flow cryostat at cryogenic temperature ($T \lesssim \SI{10}{\kelvin}$) in a time-resolved pump-probe scheme.
A mode-locked laser provides the pump pulses, while a nonlinear fiber is used to generate broadband probe pulses (see the Supplemental Material (SM) \cite{SM} for further details).
As the light shift depends on the instantaneous pump intensity at the position of the monolayer, we use a larger pump spot and few times longer pump pulse compared to the probe pulse in order to reduce averaging effects. The pump peak intensity is varied between $\SI{1}{\mega\watt\per\centi\meter\squared}$ and $\SI{100}{\mega\watt\per\centi\meter\squared}$.

\begin{figure}
\includegraphics{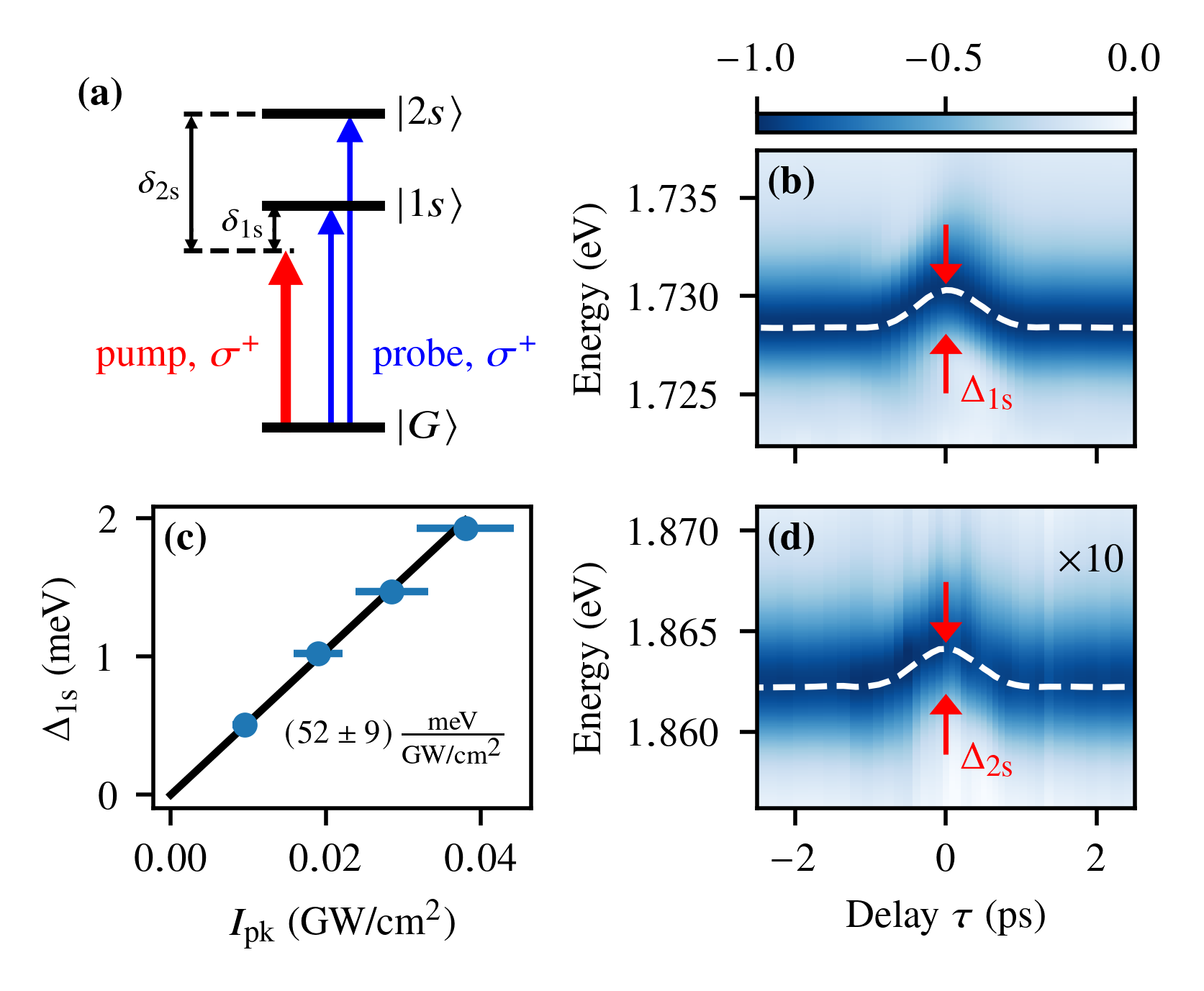}
\caption{\label{fig2} Co-circular pump-probe measurements. The pump is red detuned to the 1s state, while the probe is tuned to the 1s or 2s resonance as illustrated in the energy level diagram (a). (b),(d) Differential reflection spectra of the 1s and 2s state in WSe$_2$ as a function of the delay between pump and probe ($\delta_\text{1s}=\SI{30}{\milli\electronvolt}$, $I_\text{pk}^\text{1s} = \SI{46}{\mega\watt\per\centi\meter\squared}$, $I_\text{pk}^\text{2s} = \SI{38}{\mega\watt\per\centi\meter\squared}$). The signal in (d) is multiplied by 10 to enhance the visibility. The resonance energy for each delay step is determined by a fit and shown by the white dashed line. From this, the amplitude of the light shift $\Delta_\lambda$ is extracted and plotted against the peak intensity in (c) to extract the normalized light shift $\Delta_\lambda / I_\text{pk}$ from the slope of a linear fit.}
\end{figure}

%%%%%%%%%%%%%%%%%%%%%%%%%%%%%%%%%%%%%%%%%%%%%%%%%%%%%%%%%%%%%%%%%%%%%%%%%%%%%
%Co-circular measurements: intro to measurement method
%%%%%%%%%%%%%%%%%%%%%%%%%%%%%%%%%%%%%%%%%%%%%%%%%%%%%%%%%%%%%%%%%%%%%%%%%%%%%
\textit{Repulsive interaction.}---In a first set of experiments, we investigate the interaction of excitons within the same valley. To this end, we use co-circularly polarized pump and probe pulses with the pump pulse red detuned by $\delta_{1s}$ relative to the 1s transition (Fig.\,\ref{fig2}(a)).
At zero time delay, we observe a blueshift of the exciton resonances for both the 1s and the 2s state, which completely vanishes when pump and probe do not overlap temporally (Fig.\,\ref{fig2}(b),(d) for the example of the WSe$_2$ ML). This emphasizes the coherent nature of the induced light shift, where due to the extremely short-lived virtual excitation, relaxation and scattering processes do not play a role. During the pump pulse, the resonances broaden only marginally (typically less than $\SI{10}{\percent}$).
We fit the reflectance spectrum with a Lorentzian for each delay step and extract the amplitude of the light shift $\Delta_\lambda$ reached at zero time delay (see the SM \cite{SM} for further details). For each detuning, the light shift $\Delta_\lambda$ is measured for four different peak intensities, which exhibits the expected linear dependence on the intensity (Fig.\,\ref{fig2}(c)). Accordingly, the light shift normalized to the peak intensity $\Delta_\lambda / I_\text{pk}$ is specified throughout this work. We adjust the peak intensity so that a maximum light shift in the range of $\SI{1}{\milli\electronvolt}$ to $\SI{2}{\milli\electronvolt}$ is achieved in order to always stay within the linear regime. We observe that linearity is maintained for Stark shifts of up to $\SI{14}{\milli\electronvolt}$. 
Furthermore, the pump bandwidth is gradually reduced from $\approx\SI{7}{\milli\electronvolt}$ at large detunings to $\approx\SI{3}{\milli\electronvolt}$ for small detunings ($\delta_\text{1s}\lesssim \SI{30}{\milli\electronvolt}$).

%%%%%%%%%%%%%%%%%%%%%%%%%%%%%%%%%%%%%%%%%%%%%%%%%%%%%%%%%%%%%%%%%%%%%%%%%%%%%
%Motivation for approximation of Eq. 1
%%%%%%%%%%%%%%%%%%%%%%%%%%%%%%%%%%%%%%%%%%%%%%%%%%%%%%%%%%%%%%%%%%%%%%%%%%%%%
For our experimental parameters, we can simplify Eq.\,(\ref{eq:shift_copol}) by neglecting insignificant terms in the sum:
Since we work with a red detuning $\delta_\text{1s}$ in the range of $\SI{10}{\milli\electronvolt}$ to $\SI{100}{\milli\electronvolt}$ and the energetic splitting between the 2s and 1s resonances in WSe$_2$ and MoSe$_2$ is larger than $\SI{100}{\milli\electronvolt}$, the detuning to the Rydberg states  is always at least two times larger than the detuning to the 1s state.
%$\delta_{\lambda>1}/\delta_\text{1s}>2$ 
%Accordingly, the detuning-dependent terms $1/\delta_\lambda$ and $1/\delta_\lambda^2$ are always the largest for the 1s state. 
Furthermore, the induced polarization scales with the square root of the oscillator strength, while the virtual density scales linearly with the oscillator strength \cite{SM}. Since the oscillator strength decreases rapidly for higher Rydberg states \cite{SM}, the polarization and virtual density is mainly driven by the coupling of the pump to the 1s state, such that Eq.\,(\ref{eq:shift_copol}) can be approximated as
\begin{equation}\label{eq:shift_copol_approx}
\Delta_\lambda / I 
\approx a_{\lambda\text{1s}} \frac{1}{\delta_\text{1s}}
+ b_{\lambda \text{1s}} \frac{1}{\delta_\text{1s}^2}\quad.
\end{equation}

%%%%%%%%%%%%%%%%%%%%%%%%%%%%%%%%%%%%%%%%%%%%%%%%%%%%%%%%%%%%%%%%%%%%%%%%%%%%%
%Figure 2: Co-circular measurements
%%%%%%%%%%%%%%%%%%%%%%%%%%%%%%%%%%%%%%%%%%%%%%%%%%%%%%%%%%%%%%%%%%%%%%%%%%%%%
%%%%%%%%%%%%%%%%%%%%%%%%%%%%%%%%%%%%%%%%%%%%%%%%%%%%%%%%%%%%%%%%%%%%%%%%%%%%%
%Figure 3: Interaction strengths
%%%%%%%%%%%%%%%%%%%%%%%%%%%%%%%%%%%%%%%%%%%%%%%%%%%%%%%%%%%%%%%%%%%%%%%%%%%%%
\begin{figure}
\includegraphics{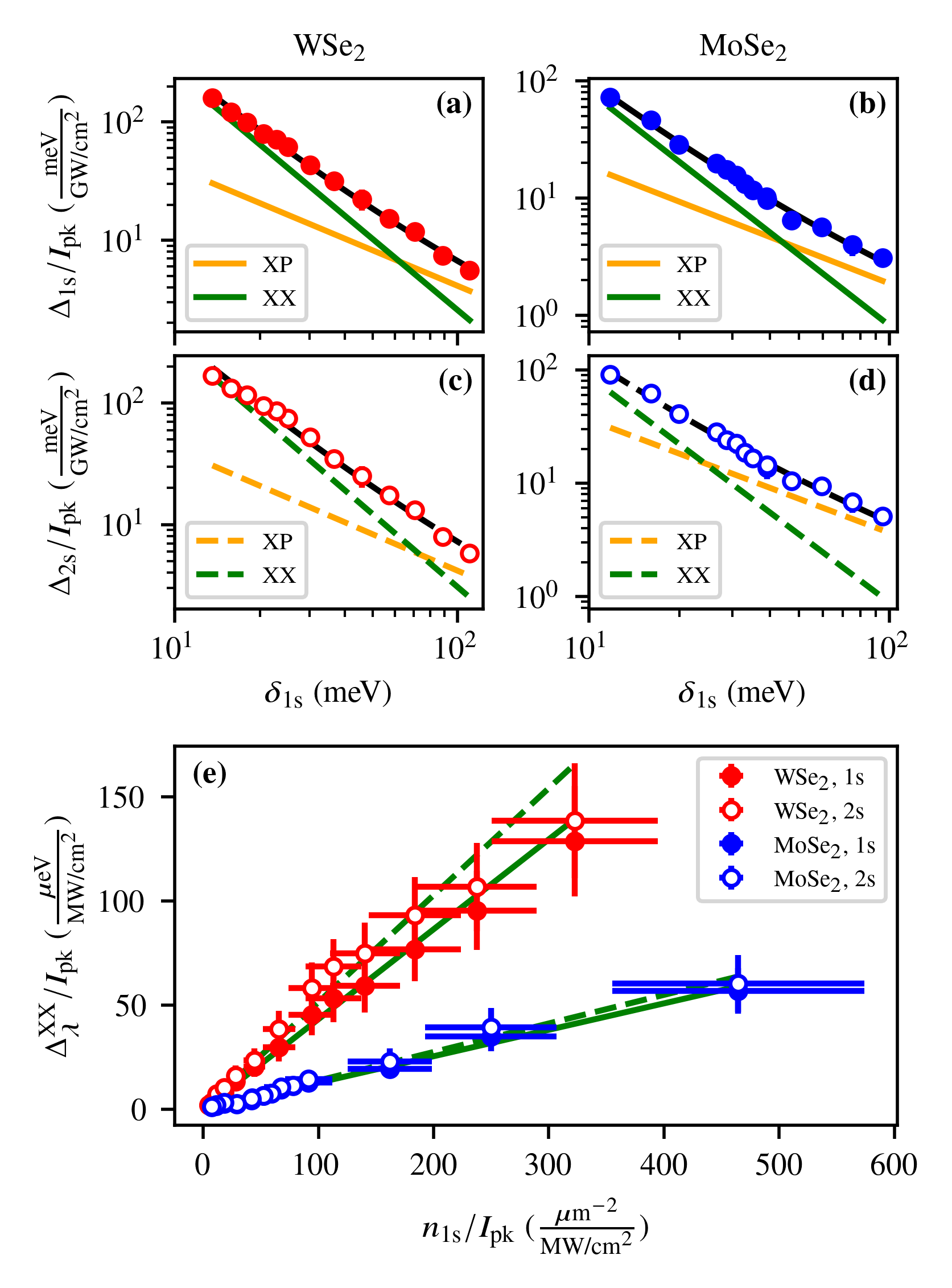}
\caption{\label{fig3} Determination of the 1s-1s and 2s-1s interaction strength. 
(a)-(d) Detuning dependence of the 1s (filled symbols) and 2s light shift (open symbols) in WSe$_2$ (red) and MoSe$_2$ (blue).
In (a)-(d) the black curve is the fit consisting of the sum of XP (orange) and XX (green) contribution as given in Eq.\,(\ref{eq:shift_copol_approx}). The XP contribution is subtracted from the data points to determine the XX interaction strength in (e), where the detuning is converted to a virtual 1s exciton density normalized to the peak intensity $n_\text{1s}/I_\text{pk}$. The XX interaction strength follows from the slope of a linear fit (green lines).}
\end{figure}
%%%%%%%%%%%%%%%%%%%%%%%%%%%%%%%%%%%%%%%%%%%%%%%%%%%%%%%%%%%%%%%%%%%%%%%%%%%%%
%1s-1s interaction
%%%%%%%%%%%%%%%%%%%%%%%%%%%%%%%%%%%%%%%%%%%%%%%%%%%%%%%%%%%%%%%%%%%%%%%%%%%%%
\textit{1s-1s interaction.}---We benchmark our measurement method by first determining the 1s-1s interaction strength.
In contrast to previous studies \cite{scuri_large_2018,uto_interaction-induced_2024}, we determine the virtual exciton density  based on the XP part of the coherent Stark shift alone.
The detuning dependence of the 1s energy shifts is well reproduced by Eq.\,(\ref{eq:shift_copol_approx}) for both materials, with the prefactors $a_\text{1s1s}$ and $b_\text{1s1s}$ as the only free fit parameters (black curves in Fig.\,\ref{fig3}(a),(b)).
Apart from the XX contribution that dominates for small detunings, we also observe a significant XP contribution, which we use to quantify the virtual exciton density $n_\text{1s}$ \cite{SM}. 
After subtracting the XP contribution from the experimental data, the XX light shift $\Delta_\lambda^\text{XX}$ can be plotted against the virtual exciton density, as shown in Fig.\,\ref{fig3}(e).
The 1s-1s XX interaction strength $U_\text{1s1s}$ finally results from a linear fit (Tab.\,\ref{tab:ratios}). 
To our knowledge, this is the first time the 1s-1s XX interaction strength in WSe$_2$ has been measured. Our value of $\SI{0.43\pm0.09}{\micro\electronvolt\micro\meter\squared}$ lies within the error bars of the result in \cite{barachati_interacting_2018} for the tungsten-based WS$_2$, which features nearly the same exciton binding energy and Bohr radius as WSe$_2$ \cite{stier_magnetooptics_2018,goryca_revealing_2019}. Our value of $\SI{0.127\pm0.029}{\micro\electronvolt\micro\meter\squared}$ for MoSe$_2$ agrees with previous studies on the same material \cite{scuri_large_2018,tan_interacting_2020, uto_interaction-induced_2024}. 
Comparing the two materials, the interaction strength in WSe$_2$ is approximately three times greater than in MoSe$_2$. This hierarchy is also theoretically expected due to the smaller Bohr radius of the excitons in MoSe$_2$ \cite{stier_magnetooptics_2018,goryca_revealing_2019}. As noted in previous studies \cite{scuri_large_2018,tan_interacting_2020, uto_interaction-induced_2024}, we find that theory \cite{shahnazaryan_exciton-exciton_2017} overestimates the interaction strength in MoSe$_2$ by a factor three, though the reason remains unclear. In WSe$_2$, the theoretical value is $\approx\SI{50}{\percent}$ greater than the experimental value.
%%%%%%%%%%%%%%%%%%%%%%%%%%%%%%%%%%%%%%%%%%%%%%%%%%%%%%%%%%%%%%%%%%%%%%%%%%%%%
%Table 1: Interaction strengths and XP/XX ratios
%%%%%%%%%%%%%%%%%%%%%%%%%%%%%%%%%%%%%%%%%%%%%%%%%%%%%%%%%%%%%%%%%%%%%%%%%%%%%
\begin{table}
\caption{\label{tab:ratios} 
1s-1s interaction strength $U_\text{1s1s}$, the 2s-1s to 1s-1s interaction strength ratio $U_\text{2s1s}/U_\text{1s1s}$ and the photonic enhancement ratio $a_\text{2s1s}/a_\text{1s1s}$ in WSe$_2$ and MoSe$_2$. The experimental values are compared with the theoretical values calculated on the basis of the SBE \cite{SM}.}
\begin{ruledtabular}
\begin{tabular}{lcc}
& WSe$_2$&MoSe$_2$\\
\hline
$U_\text{1s1s}$ exp. ($\si{\micro\electronvolt\micro\meter\squared}$) &$0.43\pm0.09$ & $0.127\pm0.029$\\ 
\vspace{-7pt}\\
$U_\text{2s1s}/U_\text{1s1s}$ exp. & $1.19\pm0.09$ & $1.08\pm0.12$\\
$U_\text{2s1s}/U_\text{1s1s}$ th. & $1.33\pm0.11$ &$ 1.40\pm0.18$\\
\vspace{-7pt}\\
$a_\text{2s1s}/a_\text{1s1s}$ exp. & $1.28\pm0.09$ & $1.77\pm0.10$\\
$a_\text{2s1s}/a_\text{1s1s}$ th. & $1.51\pm0.04$ & $1.51\pm0.18$\\
\end{tabular}
\end{ruledtabular}
\end{table}

%%%%%%%%%%%%%%%%%%%%%%%%%%%%%%%%%%%%%%%%%%%%%%%%%%%%%%%%%%%%%%%%%%%%%%%%%%%%%
%2s-1s interaction
%%%%%%%%%%%%%%%%%%%%%%%%%%%%%%%%%%%%%%%%%%%%%%%%%%%%%%%%%%%%%%%%%%%%%%%%%%%%%
\textit{2s-1s interaction.}---Adressing the 2s state with the probe, we observe a sizeable energy shift of this excited state in both materials. This shift is even larger than that of the 1s state for the same parameters of the pump, although the detuning to the 2s state is always larger than $\SI{140}{\milli\electronvolt}$.
This is because the polarization generated by the coupling of the pump to the 1s state also has a strong effect on the 2s state. 
The 1s polarization induces a dipole moment for the 2s state, which leads to an XP contribution. At the same time, the interaction between virtual 1s excitons and the 2s exciton leads to an XX contribution.
The detuning dependence is therefore analogous to that of the 1s state, as motivated in the derivation of Eq.\,(\ref{eq:shift_copol_approx}), which we fit with the parameters $a_\text{2s1s}$ and $b_\text{2s1s}$ (Fig.\,\ref{fig3}(c),(d)).

The 2s-1s interaction strength can be compared to the 1s-1s interaction strength via the ratio $b_\text{2s1s}/b_\text{1s1s} = U_\text{2s1s}/U_\text{1s1s}$. This ratio is very robust, since the interaction in both cases is due to exactly the same virtual density of 1s excitons, so that systematic errors cancel out. Since Rydberg excitons have a larger radius, one might expect a significant enhancement of the 2s-1s interaction strength.
However, our measurements show that the 2s-1s interaction is only slightly larger than the 1s-1s interaction strength ($\SI{19}{\percent}$ larger in WSe$_2$, $\SI{8}{\percent}$ larger in MoSe$_2$, see Tab.\,\ref{tab:ratios} and Fig.\,\ref{fig3}(e)).
If the Coulomb screening and the exciton wave function are taken into account in the theoretical calculations, an enhancement of $30$ to $\SI{40}{\percent}$ is obtained.
The small increase in interaction strength can be explained by several factors. Firstly, the size of the interaction is limited by the fact that the one partner in the 2s-1s interaction is still the 1s exciton. In addition, the radius of the 2s state increases $\approx\SI{40}{\percent}$ less than expected in the 2D hydrogen atom model due to Coulomb screening \cite{stier_magnetooptics_2018,goryca_revealing_2019,shahnazaryan_exciton-exciton_2017,SM}. Finally, the 2s wave function has a node, which, in addition to a dominant repulsive contribution, also results in a small attractive contribution to the interaction.

Apart from the XX contribution, we learn from the XP contribution and the ratio $a_\text{2s1s}/a_\text{1s1s}$ how much more sensitive the 2s state is to the polarization induced through the 1s state. This enhancement depends solely on the wave functions and should theoretically be around $\approx\SI{50}{\percent}$ due to the greater extent of the 2s wave function. Here we observe a clear difference between WSe$_2$ and MoSe$_2$. In WSe$_2$ the enhancement is smaller than expected with about $\SI{20}{\percent}$, while in MoSe$_2$ we observe an enhancement of almost $\SI{80}{\percent}$.

Calculating the absolute values of the exciton-exciton interaction strengths is challenging. In contrast to this, the ratio between the 2s and 1s energy shifts is well described by the SBE, regardless of the material. In addition to the 1s-1s interaction strength, our experiments provide a precise ratio of the 2s-1s to 1s-1s interaction strength for each of the materials.
A key advantage of this measurement method is the avoidance of real excitation by using a red-detuned pump (relative to the 1s state). However, this results in mainly a virtual 1s exciton density being generated, restricting the investigation to 1s-$n$s interactions.	
We tried to analyze the 2s-2s XX interaction, which is predicted to be attractive \cite{shahnazaryan_exciton-exciton_2017}, by pumping blue-detuned to the 1s resonance, in order to induce a sizeable 2s virtual exciton density. However, the 2s resonance was instantly quenched by the incoherent absorption of charge carriers, obscuring any possible coherent XX light shift. Further details on our theoretical calculations based on \cite{haug_quantum_2005,shahnazaryan_exciton-exciton_2017,slobodeniuk_semiconductor_2022} are given in the SM \cite{SM}.

\begin{figure}
\includegraphics{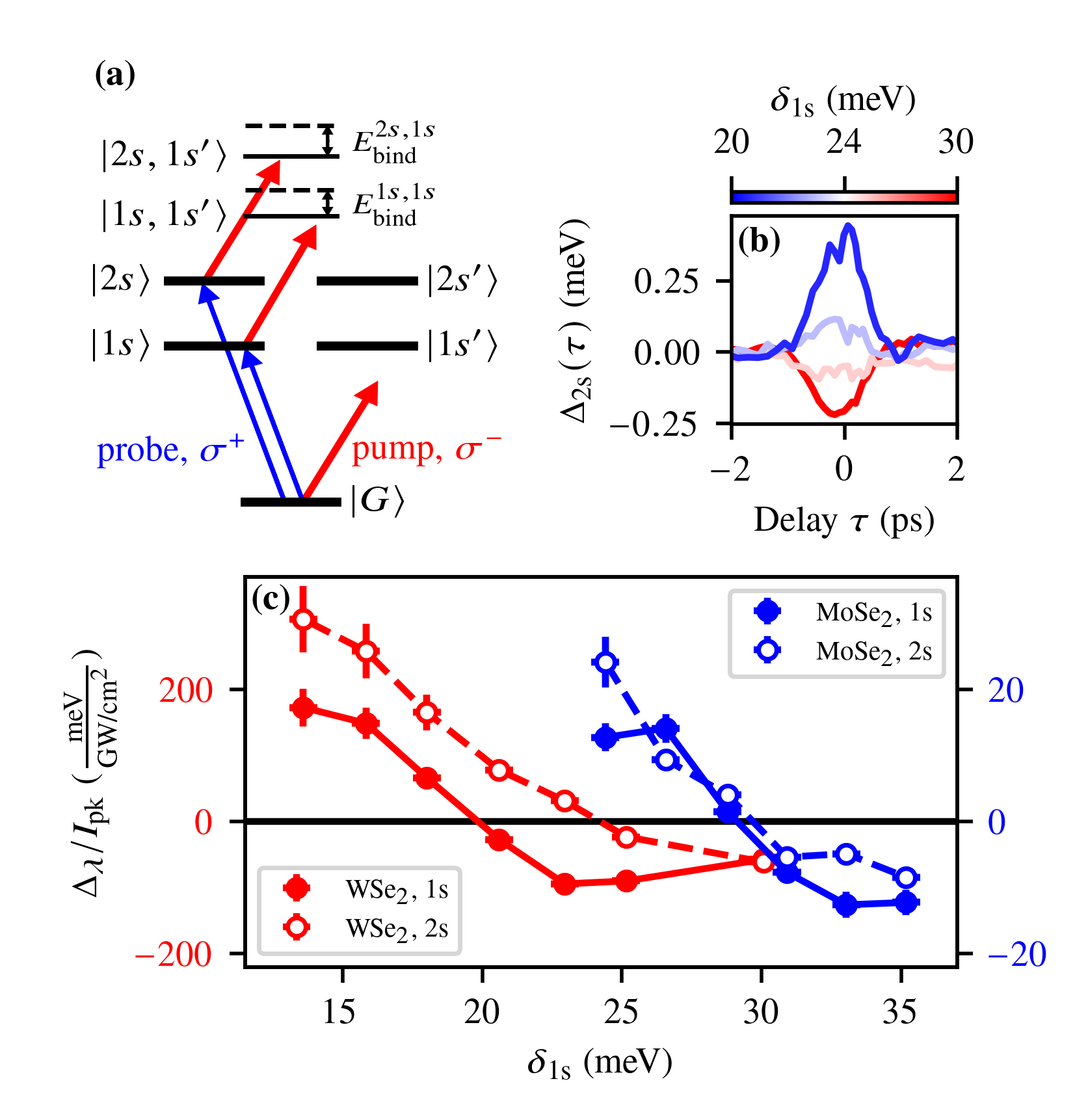}
\caption{\label{fig4}Cross-circular measurements. (a) Energy level diagram. The $\sigma^-$-polarized pump drives the transition between the $\ket{n\text{s}}$ exciton in the $K$ valley and the BX state $\ket{\text{$n$s,1s}^\prime}$. The $\sigma^+$-polarized probe monitors the energy shift of the $\ket{\text{$n$s}}$ exciton. (b) 2s light shift in WSe$_2$ as a function of the delay $\tau$ for various detunings $\delta_\text{1s}$.  (c) Light shift as a function of the detuning $\delta_\text{1s}$. The BX binding energy is extracted from the zero-crossing.}
\end{figure}

%%%%%%%%%%%%%%%%%%%%%%%%%%%%%%%%%%%%%%%%%%%%%%%%%%%%%%%%%%%%%%%%%%%%%%%%%%%%%
%Cross-circular
%%%%%%%%%%%%%%%%%%%%%%%%%%%%%%%%%%%%%%%%%%%%%%%%%%%%%%%%%%%%%%%%%%%%%%%%%%%%%
\textit{Biexcitonic state.}---While we observe the interaction between 2s and 1s excitons within the same valley to be repulsive, it is unclear whether 2s and 1s excitons from opposite valleys interact attractively and form a bound biexciton (BX) state similar to the 1s-1s BX \cite{hao_neutral_2017, ye_efficient_2018,barbone_charge-tuneable_2018,li_revealing_2018,chen_coulomb-bound_2018,steinhoff_biexciton_2018}.
To address this question, we use cross-circularly polarized pump and probe pulses, where the $\sigma^-$-polarized pump is red detuned to the 1s transition. For a detuning around the BX binding energy, the pump drives the transition between the $\ket{\text{$n$s}}$ exciton in the $K$ valley and the BX state $\ket{\text{$n$s,1s}^\prime}$ as illustrated in Fig.\,\ref{fig4}(a). 
%%%%%%%%%%%%%%%%%%%%%%%%%%%%%%%%%%%%%%%%%%%%%%%%%%%%%%%%%%%%%%%%%%%%%%%%%%%%%
%Figure 4: BX energy shift sign-change
%%%%%%%%%%%%%%%%%%%%%%%%%%%%%%%%%%%%%%%%%%%%%%%%%%%%%%%%%%%%%%%%%%%%%%%%%%%%%
The probe pulse is chosen to be on resonance with the state $\ket{\text{$n$s}}$ with $\sigma^+$-polarized light. When the pump detuning to the 1s exceeds the BX binding energy $\delta_\text{1s}>E_\text{bind}^\text{$n$s,1s}$, the optical Stark effect is expected to lead to a redshift in the probe signal. A pump detuning smaller than the BX binding energy $\delta_\text{1s}<E_\text{bind}^\text{$n$s,1s}$ results in a blueshift \cite{combescot_semiconductors_1992}. We determine the BX binding energy based on the sign change of the energy shift. To this end, we use low intensities for the pump pulse to avoid the hybridization of the dressed states at $\delta_\text{1s}=E_\text{bind}^\text{$n$s,1s}$, that causes an Autler-Townes-like splitting \cite{yong_biexcitonic_2018, SM}. We clearly identify the characteristic zero-crossing in our measurements, as shown in Fig.\,\ref{fig4}(c), which is independent of the pump intensity.
%%%%%%%%%%%%%%%%%%%%%%%%%%%%%%%%%%%%%%%%%%%%%%%%%%%%%%%%%%%%%%%%%%%%%%%%%%%%%
%Table 2: BX binding energies
%%%%%%%%%%%%%%%%%%%%%%%%%%%%%%%%%%%%%%%%%%%%%%%%%%%%%%%%%%%%%%%%%%%%%%%%%%%%%
\begin{table}
\caption{\label{tab:bind} Measured binding energies $E_\text{bind}$ of the 1s-1s and 2s-1s BX in WSe$_2$ and MoSe$_2$.}
\begin{ruledtabular}
\begin{tabular}{lcc}
$E_\text{bind} (\si{\milli\electronvolt})$& WSe$_2$&MoSe$_2$\\
\hline
1s-1s BX         & $19.9\pm0.7$    & $29.1\pm0.6$\\
2s-1s BX         & $24.2\pm1.0$    & $29.7\pm0.6$\\
%\vspace{-7pt}\\
%1s-AP$_+^s$        &$19.9\pm0.14$    &$23.10\pm0.35$\\
%1s-AP$_-^t$        &$28.37\pm0.06$   & n.a.\\
%1s-AP$_-^s$  & $35.60\pm0.08$  & $24.99\pm0.04$
\end{tabular}
\end{ruledtabular}
\end{table}

From our measurements for the 1s-1s BX (see filled data points in Fig.\,\ref{fig4}(c)) we extract binding energies (Tab.\,\ref{tab:bind}) consistent with previous studies featuring h-BN encapsulated and charge-controlled devices for both WSe$_2$ \cite{barbone_charge-tuneable_2018, ye_efficient_2018, li_revealing_2018, chen_coulomb-bound_2018} and MoSe$_2$ \cite{tan_bose_2023,uto_interaction-induced_2024}. 
It should be emphasized that we measure the inter-valley bright-bright BX, because the pump and probe only couple to the bright states and the red-detuned non-resonant excitation prevents the population of any dark states. This is in contrast to photoluminescence spectroscopy of the BX in WSe$_2$, where the emission is understood to stem from an inter-valley BX composed of a bright and a dark exciton with reported binding energies of 16 to 20 meV \cite{barbone_charge-tuneable_2018, ye_efficient_2018, li_revealing_2018, chen_coulomb-bound_2018}. 
This suggests that the binding energies of the bright-bright and bright-dark BX in WSe$_2$ are similar.

Addressing the 2s transition with the probe pulse, we observe a similar behavior of shifts (Fig.\,\ref{fig4}(b)) with a zero crossing. We interpret this as a clear signature for the first observation of a 2s-1s BX and extract binding energies of $\SI{24.2\pm1.0}{\milli\electronvolt}$ for WSe$_2$ and $\SI{29.7\pm0.6}{\milli\electronvolt}$ for MoSe$_2$.
Interestingly, in WSe$_2$ the 2s-1s binding energy is greater than the 1s-1s binding energy, while the two are almost identical in MoSe$_2$. For the repulsive interaction within the same valley we observe a similar trend when comparing the 2s-1s interaction strength with the 1s-1s interaction strength. To the best of our knowledge there is no theoretical prediction for the 2s-1s BX binding energy in these materials. 

%%%%%%%%%%%%%%%%%%%%%%%%%%%%%%%%%%%%%%%%%%%%%%%%%%%%%%%%%%%%%%%%%%%%%%%%%%%%%
%Conclusion and outlook
%%%%%%%%%%%%%%%%%%%%%%%%%%%%%%%%%%%%%%%%%%%%%%%%%%%%%%%%%%%%%%%%%%%%%%%%%%%%%
\textit{Conclusion and outlook.}---
We conducted measurements on the excitonic optical Stark effect beyond the 1s state in WSe$_2$ and MoSe$_2$ monolayers using pump-probe spectroscopy. 
The 2s exciton experiences a sizeable energy shift, despite pump detunings of more than $\SI{140}{\milli\electronvolt}$ to the 2s state. This stems from the polarization in the semiconductor induced through the interaction of the pump field with the 1s state. 
In the co-polarized measurements, we find repulsive 2s-1s interaction which is slightly larger than the 1s-1s interaction strength, in agreement with theoretical calculations based on the SBE. 
In the cross-polarized configuration, we observe a bound biexciton state of 2s and 1s excitons and determine its binding energy in both materials. These results serve as a benchmark for testing new theoretical models on exciton-exciton interactions. 
Moreover, our work opens up perspectives for the coherent manipulation of Rydberg polaritons in optical cavities \cite{walther_giant_2018, gu_enhanced_2021}.

%%%%%%%%%%%%%%%%%%%%%%%%%%%%%%%%%%%%%%%%%%%%%%%%%%%%%%%%%%%%%%%%%%%%%%%%%%%%%
%Acknowledgments
%%%%%%%%%%%%%%%%%%%%%%%%%%%%%%%%%%%%%%%%%%%%%%%%%%%%%%%%%%%%%%%%%%%%%%%%%%%%%
\textit{Acknowledgments.}---
We thank Jan-Lucas Uslu for his assistance in setting up the automatic flake-search and Susanne Zigann-Wack for her support in wire-bonding the samples.
We also thank Lutz Waldecker for insightful discussions.
We acknowledge funding from the Deutsche Forschungsgesellschaft (DFG) through the Cluster of Excellence Matter and Light for Quantum Computing (ML4Q) EXC 2004/1–390534769.

\textit{Data availability.}---
The supporting data for this article are openly available at Zenodo \cite{zenodo}.

%\bibliography{refs.bib}


\clearpage
\section*{Supplemental Material}
\section{Semiconductor Bloch equations}\label{ap:SBE}
Following the derivation of the excitonic optical Stark effect in the low-intensity regime in \cite{haug_quantum_2005,slobodeniuk_semiconductor_2022} the exciton energy shift $\Delta_\lambda$ is given by
\begin{equation}
\Delta_\lambda = \Delta_\lambda^\text{XP} + \Delta_\lambda^\text{XX}
\end{equation}
where $\Delta_\lambda^\text{XP}$ is the anharmonic interaction between the exciton and the pump field and $\Delta_\lambda^\text{XX}$ the exciton-exciton interaction:
\begin{align}
\Delta_\lambda^\text{XP} &= 2\mathcal{E}_p\sum_{k}\psi_{\lambda,k}^*d_{cv}p_k^*\psi_{\lambda,k}\label{eq:XP_theory} \\ 
\Delta_\lambda^\text{XX} &= 2 \sum_{k,k^\prime} V_{k-k^\prime}\psi_{\lambda,k}\left(p_k^*-p_{k^\prime}^*\right)\left(p_k\psi_{\lambda, k^\prime} + p_{k^\prime}\psi_{\lambda, k}\right)\quad.\label{eq:XX_theory}
\end{align}
The variable $\mathcal{E}_p$ denotes the pump field strength and $d_{cv}$ the transition dipole moment for exciting a charge carrier from the valence to the conduction band. The polarization $p_k$ induced by the pump depends on the detuning $\delta_\lambda$ and the excitonic wave functions $\psi_\lambda$:
\begin{equation}\label{eq:pk}
p_k = d_{cv}\mathcal{E}_p \sum_{\lambda} \frac{\psi_{\lambda, k}\psi_\lambda(r=0)}{\delta_\lambda}\quad.
\end{equation}
For the interaction potential $V_{k}$ we use the Rytova-Keldysh potential \cite{keldysh_l_v_coloumb_1979,cudazzo_dielectric_2011,rytova_screened_2020,slobodeniuk_semiconductor_2022}, which takes into account the screening of the Coulomb interaction due to the dielectric environment of the monolayer
\begin{equation}
V_k = \frac{e^2}{4\pi\epsilon_0}\frac{2\pi}{k\left(k r_0 + \epsilon\right)}
\end{equation}
where $r_0$ is the screening length  and $\epsilon$ the dielectric constant.

To approximate the 2D exciton wave functions we use the 2D hydrogen wave functions where the Bohr radius $r^\text{B}_\lambda$ is taken as a variational parameter to take into account the screening \cite{shahnazaryan_exciton-exciton_2017,slobodeniuk_semiconductor_2022}. The wave functions in momentum space $\psi_{\lambda, k}$ are obtained by Fourier transformation. 

The exciton oscillator strength $f_\lambda$ is given by \cite{haug_quantum_2005}
\begin{equation}\label{eq:fosc_SBE}
f_\lambda = \vert d_{cv}\vert^2 \, \vert \psi_{\lambda}(r=0) \vert^2= \vert d_{cv} \vert^2 \,   \frac{2}{\pi\,(2n-1)^3} \,\frac{1}{(r^\text{B}_\lambda)^2}
\end{equation}
and is only non-zero for the s-states $\lambda =1\text{s}, 2\text{s},\dots,n\text{s}$ with principal quantum number $n$.

We use experimentally determined values of the root-mean-square (rms) exciton radius $r_\lambda^\text{rms}$ and the screening parameters $r_0$ and $\epsilon$ measured in \cite{stier_magnetooptics_2018, goryca_revealing_2019} (Tab.\,\ref{tab:radii}). The rms exciton radius is related to the Bohr radius by 
\begin{align}
r_\text{1s}^\text{rms} &= \sqrt{\frac{3}{2}} \, r^\text{B}_\text{1s}\\
r_\text{2s}^\text{rms} &= \sqrt{\frac{117}{2}} \, r^\text{B}_\text{2s}
\end{align}
for the 1s and 2s state. We add a generous error of $\SI{10}{\percent}$ onto the Bohr radii to account for inaccuracies in our approximate wave functions. For $r_\text{2s}^\text{rms}$ in MoSe$_2$ we assume a larger error of $\SI{20}{\percent}$, since this value was not explicitly stated in \cite{goryca_revealing_2019} and is our estimate based on the information given in the SM of \cite{goryca_revealing_2019}.
\begin{table}
\caption{\label{tab:radii} The rms exciton radii $r^\text{rms}_\lambda$, screening length $r_0$ and dielectric constants $\epsilon$ taken from \cite{stier_magnetooptics_2018,goryca_revealing_2019}. The rms exciton radius of the 2s state in MoSe$_2$ was not explicitly stated in \cite{goryca_revealing_2019} and is our estimate based on the information given in the SM of \cite{goryca_revealing_2019}. From the rms radii we deduce the Bohr radii $r^\text{B}_\lambda$ and add a generous error to account for inaccuracies in our theoretical model.}
\begin{ruledtabular}
\begin{tabular}{lcc}
& WSe$_2$&MoSe$_2$\\
\hline
$r_\text{1s}^\text{rms}$  &$\SI{1.7}{\nano\meter}$ & $\SI{1.1}{\nano\meter}$\\
$r_\text{2s}^\text{rms}$  &$\SI{6.6}{\nano\meter}$ & $\approx\SI{4.5}{\nano\meter}$\\
$r_0$ & $\SI{4.5}{\nano\meter}$ & $\SI{3.9}{\nano\meter}$\\
$\epsilon$ & 4.5 & 4.4\\
\vspace{-7pt}\\
$r^\text{B}_\text{1s}$  &$\SI{1.39\pm0.14}{\nano\meter}$ & $\SI{0.90\pm0.09}{\nano\meter}$\\
$r^\text{B}_\text{2s}$  &$\SI{0.86\pm0.09}{\nano\meter}$ & $\SI{0.59\pm0.12}{\nano\meter}$\\
\end{tabular}
\end{ruledtabular}
\end{table}

\subsection{Exciton-Photon interaction}\label{ap:SBE_XP}
Plugging in the polarization $p_k$ (Eq.\,(\ref{eq:pk})) into the expression for $\Delta_\lambda^\text{XP}$ (Eq.\,(\ref{eq:XP_theory})) one obtains \cite{haug_quantum_2005}
\begin{equation}\label{eq:SBE_XP}
\Delta_\lambda^\text{XP} = 2 \vert d_{cv}\vert^2 \vert\mathcal{E}_p \vert^2
 \sum_{\lambda^\prime} \rho_{\lambda \lambda^\prime}\, 
 \frac{1}{\delta_{\lambda^\prime}}
\end{equation}
with the enhancement factor \cite{haug_quantum_2005}
\begin{align}\label{eq:SBE_XP_rho}
\rho_{\lambda \lambda^\prime} &= \psi_{\lambda^\prime}(r=0) \,\sum_{k} \psi_{\lambda^\prime,k}^* \vert \psi_{\lambda,k} \vert^2\\
&\propto \sqrt{f_\lambda} \,\,\,\sum_{k} \psi_{\lambda^\prime,k}^* \vert \psi_{\lambda,k} \vert^2\quad.
\end{align}
Eq.\,(\ref{eq:SBE_XP}) exhibits the same structure as the first term in Eq.\,(\ref{eq:shift_copol}) in the main text since $\vert\mathcal{E}_p \vert^2 \propto I$ and $a_{\lambda\lambda^\prime} \propto \rho_{\lambda \lambda^\prime}$. The factor $a_{\lambda\lambda^\prime}$ given in the main text is related to $\rho_{\lambda \lambda^\prime}$ by
\begin{equation}\label{eq:arho}
a_{\lambda \lambda^\prime} = \rho_{\lambda \lambda^\prime} \, \frac{4 \vert d_{cv}\vert^2}{n c_0 \epsilon_0}
\end{equation}
where $n$ denotes the refractive index, $c_0$ the speed of light in vacuum and $\epsilon_0$ the vacuum permittivity.
The theoretically calculated values of $\rho_\text{1s1s}$, $\rho_\text{2s1s}$ and $\rho_\text{2s1s}/\rho_\text{1s1s} = a_\text{2s1s}/a_\text{1s1s}$ are summarized in Tab.\,\ref{tab:XPXXtheory}.

\subsection{Exciton-Exciton interaction}\label{ap:SBE_XX}
Proceeding in the same way with Eq.\,(\ref{eq:XX_theory}) for the XX interaction, we obtain the product of several sums over the excitonic states $\lambda$ due to the polarization $p_k$. Accordingly, the structure of the second term in Eq.\,(\ref{eq:shift_copol}) in the main text is already a simplified version of Eq.\,(\ref{eq:XX_theory}). 
We further approximate Eq.\,(\ref{eq:XX_theory}) for $\delta_\text{1s}\ll \delta_{n\text{s}},\, n>1$ such that $p_k \approx d_{cv}\mathcal{E}_p \, \psi_{{1}, k}\psi_\text{1s}(r=0)/\delta_\text{1s}$ yielding
\begin{equation}
\Delta_\lambda^\text{XX} = n_\text{1s} \, U_{\lambda\text{1s}}
\end{equation}
with the virtual exciton density \cite{schmitt-rink_linear_1989}
\begin{align}
n_\text{1s} &= 2 \sum_{k} \vert p_k \vert ^2 \nonumber \\
&=2 \vert d_{cv}\vert^2 \vert\mathcal{E}_p \vert^2 \, \vert  \psi_\text{1s}(r=0) \vert^2 \, \frac{1}{\delta_\text{1s}^2} \nonumber \\
&= 2 \vert\mathcal{E}_p \vert^2 \, f_\text{1s}  \,
 \frac{1}{\delta_\text{1s}^2} \label{eq:nex_SBE}
\end{align}
and the interaction strength \cite{haug_quantum_2005,slobodeniuk_semiconductor_2022}
\begin{align}
U_{\lambda\text{1s}} =& \sum_{k,k^\prime} V_{k-k^\prime}\psi_{\lambda, k}^*
\left( \psi_{\text{1s}, k}^* -\psi_{\text{1s}, k^\prime}^*      \right) \nonumber \\
&\times \left( \psi_{\text{1s}, k} \psi_{\lambda, k^\prime} + \psi_{\text{1s}, k^\prime} \psi_{\lambda, k}     \right)\quad .
\end{align}
The factor $b_{\lambda\text{1s}}$ given in the main text is related to $U_{\lambda \text{1s}}$ by
\begin{equation}
b_{\lambda\text{1s}} = U_{\lambda \text{1s}} \, \frac{4 \vert d_{cv}\vert^2 \vert  \psi_\text{1s}(r=0) \vert^2 }{n c_0 \epsilon_0}\quad.
\end{equation}
The theoretically calculated values of $U_\text{1s1s}$, $U_\text{2s1s}$ and $U_\text{2s1s}/U_\text{1s1s} = b_\text{2s1s}/b_\text{1s1s}$ are summarized in Tab.\,\ref{tab:XPXXtheory}.
\begin{table}
\caption{\label{tab:XPXXtheory} Theoretically calculated values for the XP enhancement factor $\rho_{\lambda\lambda^\prime}$ and XX interaction strengths $U_{\lambda\lambda^\prime}$ using the parameters given in Tab.\,\ref{tab:radii}.}
\begin{ruledtabular}
\begin{tabular}{lcc}
& WSe$_2$&MoSe$_2$\\
\hline
$\rho_\text{1s1s}$ & $16/7$ & $16/7$ \\
$\rho_\text{2s1s}$ & $3.45\pm0.09$ & $3.5\pm0.4$ \\
$\rho_\text{2s1s}/\rho_\text{1s1s}$ & $ 1.51\pm0.04$ & $1.51\pm0.18$ \\
\vspace{-7pt}\\
$U_\text{1s1s}$ ($\si{\micro\electronvolt\micro\meter\squared}$)& $1.19\pm0.16$ & $0.66\pm0.10$ \\
$U_\text{2s1s}$ ($\si{\micro\electronvolt\micro\meter\squared}$)& $1.58\pm0.15$ & $0.91\pm0.15$ \\
$U_\text{2s1s}/U_\text{1s1s}$ & $ 1.33\pm0.11$ & $1.40\pm0.18$
\end{tabular}
\end{ruledtabular}
\end{table}

Furthermore, the fact that $p_k$ is directly related to the virtual exciton density $n_\text{1s}$ means that we can use the XP contribution to quantify the virtual exciton density. Namely, we extract the transition dipole moment $d_{cv}$ from the XP contribution based on the fitted parameter $a_\text{1s1s}$ (Eq.\,(\ref{eq:arho})). From the XP contribution of the 1s light shift we obtain $(5.25\pm0.17)$ Debye for WSe$_2$ and $(3.54\pm0.16)$ Debye for MoSe$_2$. Together with the Bohr radii measured in \cite{stier_magnetooptics_2018,goryca_revealing_2019} $n_\text{1s}$ can be determined.

\section{Device fabrication}
\begin{figure}
\includegraphics{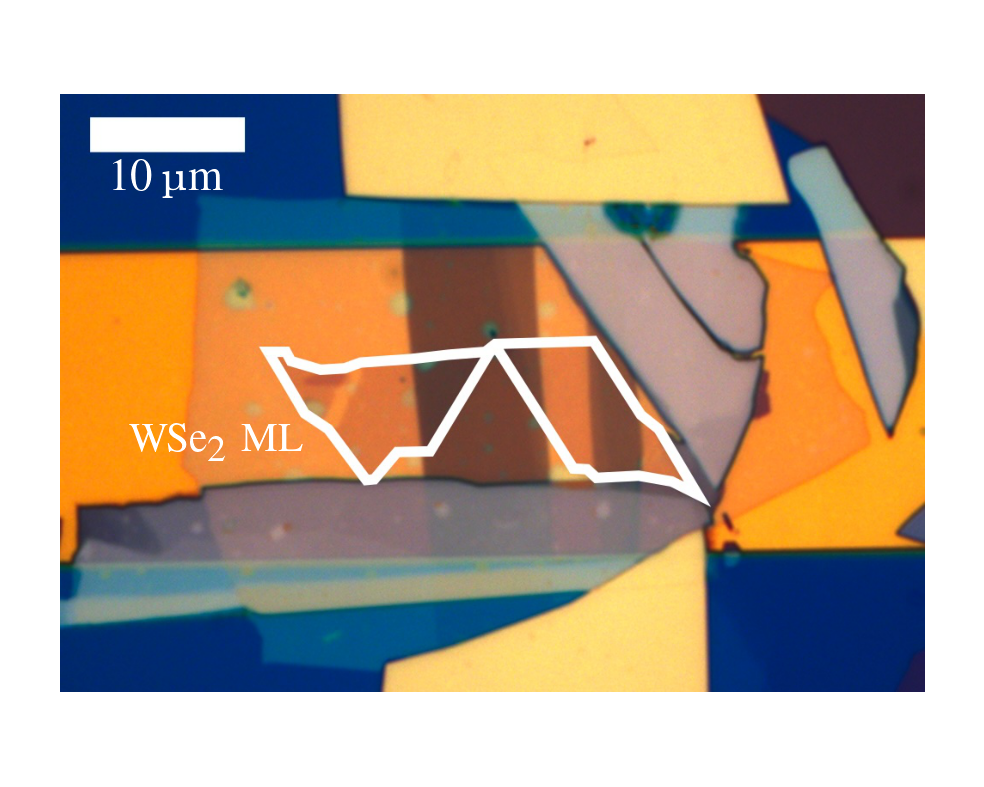}
\caption{\label{fig:wse2_device} Optical micrograph of the WSe$_2$ device.}
\end{figure}

\begin{table}
\caption{\label{tab:flakes_suppliers} Suppliers of the bulk crystals used for the mechanical exfoliation of the flakes.}
\begin{ruledtabular}
\begin{tabular}{ll}
Material & Supplier\\
\hline
h-BN & 2D semiconductors\\
graphite & NGS Naturgraphit \\
WSe$_2$ & 2D semiconductors\\
MoSe$_2$ & HQ Graphene\\
\end{tabular}
\end{ruledtabular}
\end{table}
The bulk crystals (see Tab.\,\ref{tab:flakes_suppliers}) were exfoliated onto Si/SiO$_2$ ($\SI{90}{\nano\meter}$) wafers and automatically searched with a setup and algorithm as described in \cite{uslu_open-source_2024}. The flakes are stacked using a dry-transfer technique \cite{zomer_fast_2014} employing a polycarbonate (PC) film (HQ Graphene) on top of a polydimethylsiloxane (PDMS) dome. The full stack is then deposited onto a Si/SiO$_2$ ($\SI{285}{\nano\meter}$) substrate with a pre-patterned gold backgate. The pre-patterned gold structures were fabricated via electron-beam lithography and thermal gold evaporation with a thin chromium layer between the gold and the substrate to improve the adhesion ($\SI{5}{\nano\meter}$ chromium and $\SI{80}{\nano\meter}$ gold). The TMDC monolayer is contacted using a graphite flake. The contact to the graphite flake is established with a second electron-beam lithography and gold evaporation. The gold contacts on the substrate are connected to a PCB chip carrier by wire bonding. An optical micrograph of the WSe$_2$ device is shown in Fig.\,\ref{fig:wse2_device}.
\section{Device geometries and transfer matrix simulations}\label{ap:device_tmm}
The structure of our devices is shown in Fig.\,\ref{fig1}(a) in the main text. The layer thicknesses are summarized in Tab.\,\ref{tab:stackthick}. References for the refractive indices used are also given in Tab.\,\ref{tab:stackthick}. For the TMDC monolayers, we use a constant background refractive index of $5\pm0.5$ for WSe$_2$ and $4\pm0.5$ for MoSe$_2$ \cite{liu_optical_2014}. The thicknesses of the h-BN and gold layers were measured using atomic force microscopy. The reflectance of the device is modeled based on a transfer matrix simulation, where the exciton resonance is described within the framework of the optical Bloch equations (OBE), as described in more detail in \cite{scuri_large_2018, zhou_controlling_2020, wild_algorithms_2020}.  Based on this, the radiative and non-radiative decay rates $\gamma_r$ and $\gamma_{nr}$ can be determined from the linear reflection spectrum of the device. The Purcell factor is also obtained from the transfer matrix simulation, so that the radiative rate in vacuum $\gamma_{r,0}$ can be determined. Similarly, the  field enhancement factor $f_\text{FEF} = \vert \mathcal{E}^\text{ML} \vert/\vert\mathcal{E}^\text{in}\vert$, describing how the electric field amplitude of an incoming wave $\mathcal{E}^\text{in}$ is related to the electric field amplitude at the monolayer $\mathcal{E}^\text{ML}$, is determined based on the transfer matrix simulation. The Purcell factor, field enhancement factor and decay rates are summarized in Tab.\,\ref{tab:stackparams}.
\begin{table}
\caption{\label{tab:stackthick} Layer-thicknesses of the WSe$_2$ and MoSe$_2$ devices. The references used for the refractive index of the materials are given in the first column for each material.}
\begin{ruledtabular}
\begin{tabular}{lcc}
Layer & \multicolumn{2}{c}{Thickness ($\si{\nano\meter}$)}\\
 & WSe$_2$ device&MoSe$_2$ device\\
\hline
vacuum & semi-infinite &  semi-infinite\\
h-BN (top), \cite{lee_refractive_2019} & $12\pm2$&$13\pm2$\\
TMDC, \cite{liu_optical_2014} & $0.7\pm0.2$ & $0.7\pm0.2$\\
h-BN (bottom), \cite{lee_refractive_2019} & $19\pm2$&$20\pm2$\\
gold, \cite{mcpeak_plasmonic_2015} & $86\pm2$&$86\pm2$\\
SiO$_2$, \cite{malitson_interspecimen_1965} & $285$ & $285$\\
silicon, \cite{schinke_uncertainty_2015} & semi-infinite &  semi-infinite
\end{tabular}
\end{ruledtabular}
\end{table}

\begin{table}
\caption{\label{tab:stackparams} Purcell factor and field enhancement factor $f_\text{FEF}$ calculated on the basis of a transfer matrix simulation (evaluated at the wavelength of the respective 1s resonance). The radiative decay rate $\gamma_{r,0}$ in vacuum, radiative decay rate $\gamma_{r}$ within the stack and non-radiative decay rate $\gamma_{nr}$ are extracted from the linear reflection spectrum of the WSe$_2$ and MoSe$_2$ devices.}
\begin{ruledtabular}
\begin{tabular}{lcc}
 & WSe$_2$ device&MoSe$_2$ device\\
\hline
Purcell factor &$0.59\pm0.08$ &  $0.50\pm0.07$\\
$\vert f_\text{FEF} \vert$ & $1.08\pm0.09$& $1.00\pm0.08$ \\
$\gamma_{r,0}$ & $\SI{4.4\pm0.6}{\milli\electronvolt}$ & $\SI{4.3\pm0.6}{\milli\electronvolt}$\\
$\gamma_{r}$ & $\SI{2.562\pm0.010}{\milli\electronvolt}$&$\SI{2.154\pm0.007}{\milli\electronvolt}$\\
$\gamma_{nr}$ & $\SI{2.918\pm0.009}{\milli\electronvolt}$&$\SI{2.180\pm0.006}{\milli\electronvolt}$\\
\end{tabular}
\end{ruledtabular}
\end{table}
\subsection{OBE: Virtual exciton density}\label{ap:nex_OBE}
It is also possible to determine the virtual exciton density, and subsequently the interaction strength, in the framework of the OBE. This approach was used in \cite{scuri_large_2018,uto_interaction-induced_2024,evrard_ac_2025}.
We solve the OBE given in \cite{scuri_large_2018,zhou_controlling_2020,wild_algorithms_2020} in steady state and obtain:
\begin{equation}\label{eq:nex_OBE}
n_\text{1s}^\text{OBE} = 2 \, I_{f} \, \frac{\gamma_{r,0}}{\delta_\text{1s}^2 + (\frac{\gamma_r + \gamma_{nr}}{2})^2} \approx 2 \, I_{f} \, \gamma_{r,0} \, \frac{1}{\delta_\text{1s}^2}
\end{equation}
where $I_f$ denotes the pump photon flux at the monolayer and $\gamma_{r,0}$, $\gamma_r$ and $\gamma_{nr}$ the radiative and non-radiative decay rates as discussed in the previous section and summarized in Tab.\,\ref{tab:stackparams}.

Considering that the oscillator strength is proportional to the radiative rate $\gamma_{r,0}\propto f_\text{1s}$ the formula for the virtual density within the OBE (Eq.\,(\ref{eq:nex_OBE})) has the same structure as the one within the framework of the SBE (Eq.\,(\ref{eq:nex_SBE})). Namely, the virtual  density for a given detuning $\delta_{1s}$ is proportional to the intensity of the pump multiplied by the oscillator strength.

When we calculate the virtual density using Eq.\,(\ref{eq:nex_OBE}) (OBE), we obtain a virtual density that is approximately 3 times greater than the one calculated using Eq.\,(\ref{eq:nex_SBE}) (SBE). Accordingly, the 1s-1s interaction strength extracted based on the OBE is a factor of $\approx3$ smaller. However, both methods yield the same ratio of 1s-1s interaction strengths of WSe$_2$ and MoSe$_2$ $U_\text{1s1s}^\text{WSe$_2$}/_\text{1s1s}^\text{MoSe$_2$}$. This is due to the fact, that the vacuum radiative rate $f_\text{1s}\propto \gamma_{r,0}$ is the same for both materials (Tab.\,\ref{tab:stackparams}), just as the oscillator strength $f_\text{1s}\propto \vert d_{cv} \vert^2 / (r_\text{1s}^\text{B})^2$ within the SBE (Eq.\,\ref{eq:fosc_SBE}) is the same for both materials.

\section{Experimental setup}
\begin{figure}
\includegraphics[width=8.5cm]{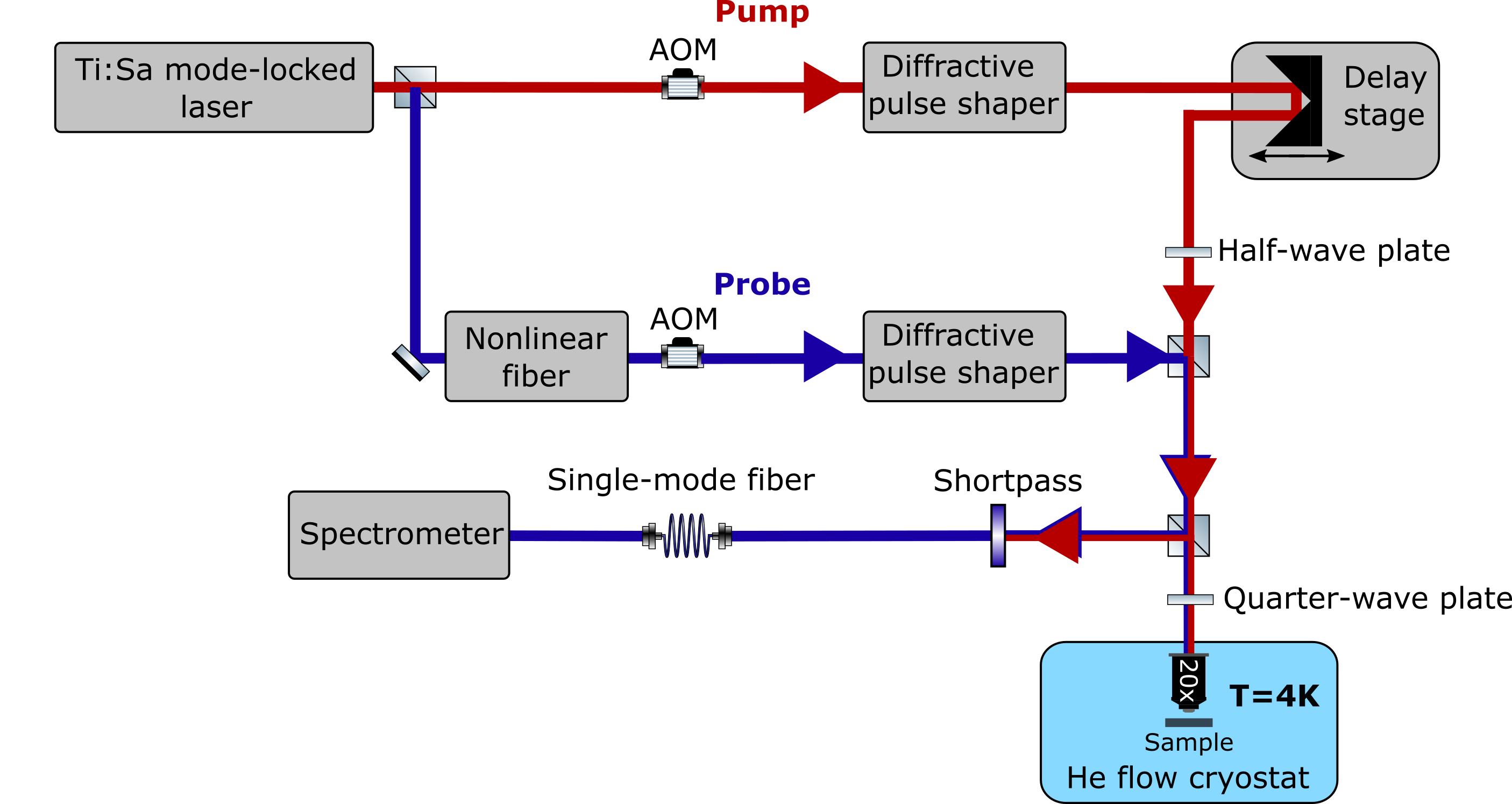}
\caption{\label{setup} Sketch of the confocal pump-probe setup in reflection geometry.}
\end{figure}

A sketch of the confocal pump-probe setup in reflection geometry is shown in Fig.\,\ref{setup}. 
A mode-locked Ti:Sa laser produces $\approx \SI{150}{\femto\second}$ pulses at a repetition rate of $\SI{80}{\mega\hertz}$ (Coherent, Chameleon Ultra 2). The laser beam is split into a pump and probe beam using a beam splitter. The delay between the pump and probe pulses is controlled through a delay stage in the pump arm, before recombining the pulses with another beam splitter. For the probe a nonlinear photonic crystal fiber is used to generate broadband pulses (NKT photonics, femtowhite 800).  The bandwidth of the pump and probe pulses are both adjusted using diffractive pulse shapers. The bandwidth of the pump pulses is reduced to a wavelength interval of around $\SI{1}{\nano\metre}$ to $\SI{3}{\nano\metre}$ yielding pulse durations of $\SI{500}{\femto\second}$ to $\SI{1000}{\femto\second}$, whereas the probe pulses feature a bandwidth of $\SI{30}{\nano\meter}$ with a pulse duration of $\SI{100}{\femto\second}$ to $\SI{300}{\femto\second}$. The pulse durations are measured using an interferometric autocorrelator.  The power of the pump and probe pulses are each adjusted using an acousto-optic modulator (AOM). An average power of $\leq \SI{100}{\nano\watt}$ reaching the sample was used for the probe, whereas an average power in the range of $\SI{10}{\micro\watt}$ to $\SI{1000}{\micro\watt}$ was typically used for the pump resulting in peak intensities on the order of $\SI{1}{\mega\watt\per\centi\meter\squared}$ to $\SI{100}{\mega\watt\per\centi\meter\squared}$. A quarter-wave plate in front of the cryostat is used to polarize pump and probe circularly. While the circular polarization of the probe is kept fixed, a half-wave plate in the pump arm is used to switch between left- and right-hand circularly polarized light. The sample is mounted onto a cold finger reaching $\approx \SI{4}{\kelvin}$ in a custom Helium-flow cryostat (Cryovac, Konti Micro). The $20\times$ microscope objective (Olympus, UPLFLN 20X) with NA=$0.5$ is located within the cryostat (but not actively cooled) and separated from the sample by a $\SI{170}{\micro\meter}$ thick glass window attached to the heat shield which is cooled by the exhaust gas. The probe pulses are focused to a diffraction-limited spot with a $1/e^2$ radius of $\SI{0.6}{\micro\meter}$, while the pump pulses feature an approximately three times larger spot size. The reflected probe is coupled into a single-mode fiber realizing a confocal setup. The probe spectrum is measured using a spectrometer and CCD camera (Teledyne Princeton Instruments, SpectraPro HRS-500-MS and Blaze 400-HR LD). The reflected pump is blocked using a shortpass filter.

\section{Calculation of the peak intensity}\label{ap:peakint}

 The peak intensity specified in the main text denotes the peak intensity at the monolayer
%, which is effectively \enquote{seen} by the probe 
and includes several correction factors taking into account the parameters of the pump and probe pulses and the geometry of the stack.
For each measurement, the pulse durations as well as the $1/e^2$ beam radii in both x- and y-direction are measured for the pump and probe. The beam radii are measured using a knife-edge scan over a gold edge on the sample. The average power of the pump is measured and converted to a peak intensity incident on the stack by taking into account the repetition rate $f_\text{rep}$, pulse duration $\tau_\text{pump}$ and beam radii $\omega_\text{x,pump}$ and $\omega_\text{y,pump}$
\begin{equation}
I_\text{pk}^\text{in} = 
\frac{P_\text{avg}}{f_\text{rep}}\,
\frac{2}{\pi\,\omega_\text{x,pump}\,\omega_\text{y,pump}}
\,\frac{f_\text{pk}}{\tau_\text{pump}}
\end{equation}
where $f_\text{pk}$ is a prefactor that depends on the temporal shape of the pump pulse. To obtain the peak intensity at the location of the monolayer the square of the field enhancement factor $f_\text{FEF}$ (Tab.\,\ref{tab:stackparams}) is multiplied with the peak intensity yielding:
\begin{equation}
I_\text{pk}^\text{ML} = \vert f_\text{FEF} \vert^2 \, I_\text{pk}^\text{in} \quad.
\end{equation}
The field enhancement factor follows from the transfer matrix simulation of the device, evaluated at the respective pump wavelength, as discussed in the previous section. Here, we note, that we do not take into account the refractive index at the position of the monolayer. Accordingly, we also set the refractive index to $n=1$ when we convert the peak intensity $I_\text{pk}$ stated in the main text back to an electric field strength $\mathcal{E}_p$ to for example determine the transition dipole moment as given in Eq.\,(\ref{eq:arho}). 

Finally, we consider the temporal and spatial extent of the probe pulse which averages in time and space over the pump intensity. Assuming Gaussian pulses in space and time we obtain two additional correction factors:
\begin{align}
f_\text{spat} &= \frac{1}{\sqrt{1+\left(\frac{\omega_\text{x,probe}}{\omega_\text{x,pump}}\right)^2}\sqrt{1+\left(\frac{\omega_\text{y,probe}}{\omega_\text{y,pump}}\right)^2}}\\
f_\text{temp} &= \frac{1}{\sqrt{1+\left(\frac{\tau_\text{probe}}{\tau_\text{pump}}\right)^2}} \label{eq:corr_temp}\quad.
\end{align}
For an infinitely small and short probe pulse, both correction factors approach $1$ as expected. All in all, this yields the peak intensity $I_\text{pk}$ as stated throughout the main text:
\begin{align}
I_\text{pk} = f_\text{spat}\, f_\text{temp}\, I_\text{pk}^\text{ML} \quad.
\end{align}
We verified Eq.\,(\ref{eq:corr_temp}) by changing the pulse duration ratio $\tau_\text{pump}/\tau_\text{probe}$ from one to five and measuring the light shift $\Delta$. When applying the correction factor the normalized light shift $\Delta/I_\text{pk}$ stays constant, validating our approach.

\section{Measurement and analysis workflow}
\subsection{Extraction of the optical Stark shift}
To measure the optical Stark shift for a given detuning and peak intensity we perform a pump-probe measurement as shown in Fig.\,\ref{fig2}(b),(d) in the main text.

First, the reflection spectrum of the exciton resonance in the absence of the pump is fitted with the function $F(E)$ with the fit parameters $A,E_0,\gamma,\theta, a, b$
\begin{align}
F(E) &= A\left[\cos{\theta}\,L_1(E) + \sin{\theta}\,L_2(E)\right]+B(E)
\end{align}
which consists of a sum of a Lorentzian $L_1(E;E_0,\gamma)$ and a dispersive Lorentzian $L_2(E;E_0,\gamma)$, and a linear background term $B(E;a,b)$:
\begin{align}
L_1(E;E_0,\gamma)&=\frac{\gamma/2}{(E-E_0)^2+(\gamma/2)^2}\\
L_2(E;E_0,\gamma)&=\frac{E-E_0}{(E-E_0)^2+(\gamma/2)^2}\\
B(E;a,b)&=a+bE \quad.
\end{align}
Next, the reflection spectrum of the exciton resonance in the presence of the pump is fitted for each delay step, where $\theta, a, b$ are kept fixed from the previous fit of the reflection spectrum in the absence of the pump. 
As a result we obtain the resonance energy $E_0^\prime(\tau)$, width $\gamma^\prime(\tau)$ and amplitude $A^\prime(\tau)$ as a function of the delay $\tau$. The plot of $E_0^\prime(\tau)$ is shown by the white dashed line in Fig.\,\ref{fig2}(b),(d) in the main text. The optical Stark shift as a function of the delay follows as $\Delta_\lambda(\tau)=E_0^\prime(\tau) - E_0$. Finally, we fit $\Delta_\lambda(\tau)$ with a Gaussian function to extract the amplitude of the Stark shift $\Delta_\lambda$ at zero time delay.

\begin{figure}
\includegraphics{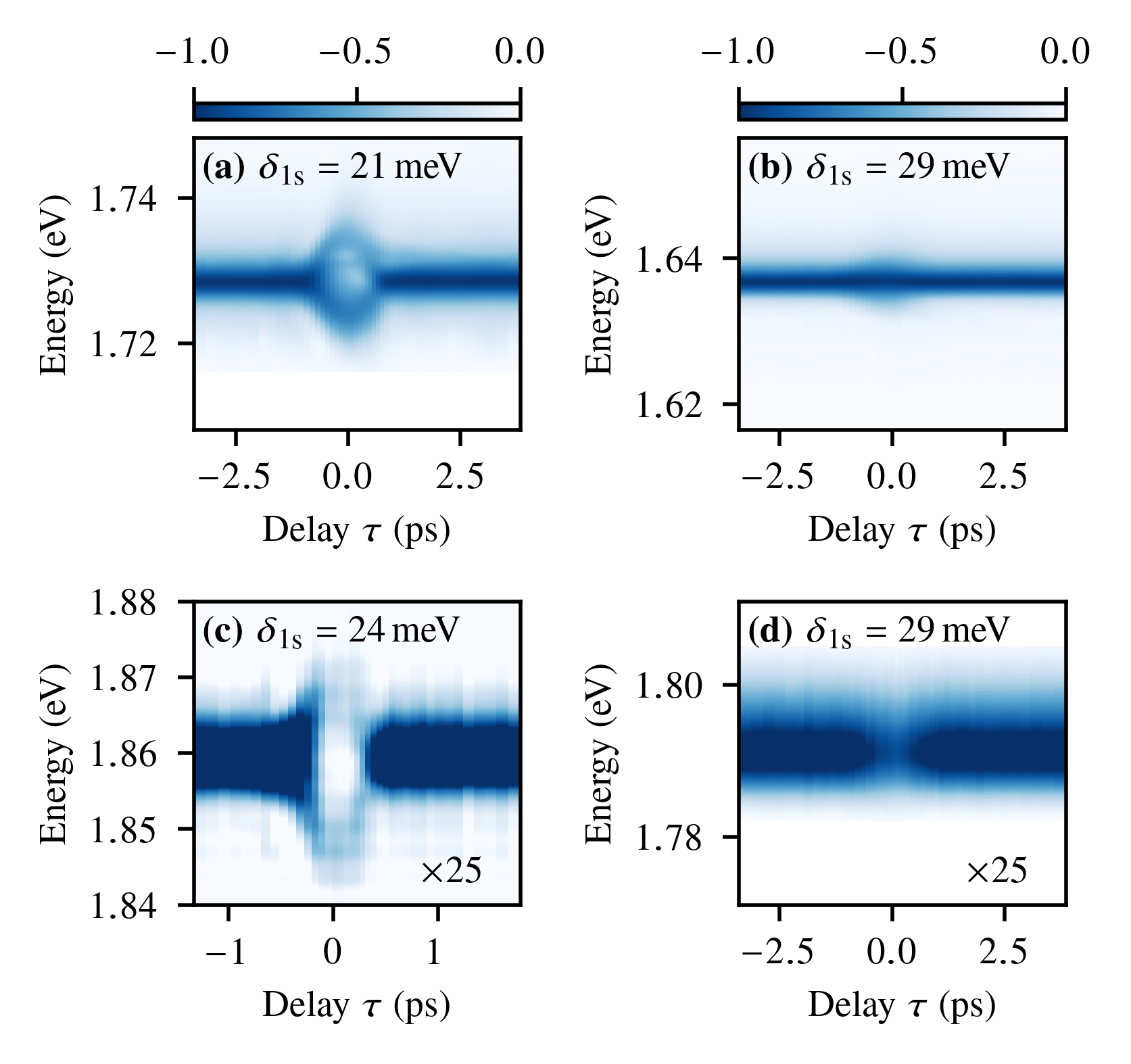}
\caption{\label{fig:BX_ATsplit} Cross-circular measurements in resonance to the BX at higher peak intensity. (a), (b) 1s resonance of WSe$_2$ and MoSe$_2$ with $I_\text{pk}=\SI{33}{\mega\watt\per\centi\meter\squared}$ and $I_\text{pk}=\SI{19}{\mega\watt\per\centi\meter\squared}$. (c), (d) 2s resonance of WSe$_2$ and MoSe$_2$ with $I_\text{pk}=\SI{52}{\mega\watt\per\centi\meter\squared}$ and $I_\text{pk}=\SI{20}{\mega\watt\per\centi\meter\squared}$. The signals are multiplied by 25 to enhance the visibility.}
\end{figure}

\subsection{Optical Stark shift as a function of the peak intensity}
While keeping the detuning fixed, the pump-probe measurements are repeated for various intensities. The maximum intensity is adjusted for each detuning such that a maximum Stark shift of $\SI{1}{\milli\electronvolt}$ to $\SI{2}{\milli\electronvolt}$ is achieved. This yields the optical Stark shift as a function of the peak intensity as shown in Fig.\,\ref{fig2}(c) in the main text. Based on this, the normalized Stark shift $\Delta_\lambda/I_\text{pk}$ is determined from a linear fit (without intercept) as shown by the black line in Fig.\,\ref{fig2}(c) in the main text. This enables us to achieve a high signal-to-noise-ratio in the determination of the normalized light shift regardless of the detuning. At the same time, we check for each detuning that we are indeed in the linear regime.
\vspace{0cm}
\subsection{Optical Stark shift as a function of the detuning: Disentangling the XP and XX contribution}
Finally, the normalized light shift $\Delta_\lambda/I_\text{pk}$ is measured for different detunings to obtain the data as shown in Fig.\,\ref{fig3}(a)-(d) in the main text. The measurements for the 1s and 2s states are carried out directly one after the other to make sure that the parameters of the pump are identical (opposed to measuring the 1s and 2s light shift in two separate measurement series). The detuning is varied over one order of magnitude and results in normalized light shifts that vary over two orders of magnitude. Finally, Eq.\,(\ref{eq:shift_copol_approx}) in the main text is fitted to these data with only two free fitting parameters, namely the amplitude of the XP and XX contribution $a_{\lambda\text{1s}}$ and $b_{\lambda\text{1s}}$, where the slope for each contribution is fixed by theory ($1/\delta_\text{1s}$ for XP and $1/\delta_\text{1s}^2$ for XX, corresponding to a fixed slope in the double-logarithmic plot). This makes the fitting extremely robust and results in relative errors of less than $\SI{10}{\percent}$ on the fit parameters.

\section{Biexcitonic Autler-Townes-like splitting}\label{ap:BX_split}
For WSe$_2$ we observe a clear Autler-Townes-like splitting as in \cite{yong_biexcitonic_2018} for higher peak intensities for both the 1s and 2s resonance (Fig.\,\ref{fig:BX_ATsplit}(a),(c)).
In the case of MoSe$_2$, however, we only observe a broadening of the resonance and no splitting (Fig.\,\ref{fig:BX_ATsplit}(b),(d)), consistent with the results in \cite{uto_interaction-induced_2024}. Comparing the normalized light shift achieved in the cross-polarized measurements, the light shift in WSe$_2$ is an order of magnitude larger than in MoSe$_2$ (Fig.\,\ref{fig4}(c) in the main text). Accordingly, a significantly higher power may simply be required to achieve the Autler-Townes splitting in MoSe$_2$, which was not available to us in this measurement.
\end{document}